\documentclass[useAMS,usenatbib]{mn2e}
\usepackage{times}
\usepackage{graphicx}
\newcommand{\etal  }{{et al.} }

\title[Collapse and Fragmentation of Magnetized Molecular Cloud Cores]
{
First MHD Simulation of Collapse and Fragmentation of Magnetized Molecular Cloud Cores
}
\author[M. ~N. ~Machida, K. ~Tomisaka and T. ~Matsumoto]
{Masahiro ~N. ~Machida$^{1}$\thanks{E-mail:machida@th.nao.ac.jp}, 
Kohji ~Tomisaka$^{1}$\thanks{E-mail:tomisaka@th.nao.ac.jp} 
and Tomoaki ~Matsumoto$^{1,2}$\thanks{E-mail:matsu@i.hosei.ac.jp}\\
$^{1}$ National Astronomical Observatory of Japan, Mitaka, Tokyo 181-8588, Japan \\
$^{2}$ Faculty of Humanity and Environment, Hosei University, Fujimi, Chiyoda-ku, Tokyo 102-8160, Japan}

\begin{document}



\maketitle

\begin{abstract}
This is the first paper about the fragmentation and mass outflow in the molecular cloud by using  three-dimensional MHD nested-grid simulations.
The binary star formation process is studied paying particular attention to the fragmentation of a rotating magnetized molecular cloud.
We assume an isothermal rotating and magnetized cylindrical cloud in hydrostatic balance.
Non-axisymmetric as well as axisymmetric perturbations are added to the initial state and the subsequent evolutions are studied.
The evolution is characterized by three parameters: the amplitude of the non-axisymmetric perturbations, the rotation speed, and the magnetic field strength.
As a result, it is found that non-axisymmetry hardly evolves in the early phase, but begins to grow after the gas contracts and forms a thin disk.
Disk formation is strongly promoted by the rotation speed and the magnetic field strength.
There are two types of fragmentation: fragmentation from a ring and that from a bar.
Thin adiabatic cores fragment if a thickness is smaller than 1/4 of the radius.
For the fragments to survive, they should be formed in a heavily elongated barred core or a flat round disk. 
In the models showing fragmentation, outflows from respective fragments are found as well as those driven by the rotating bar or the disk. 
\end{abstract}


\section{Introduction}
	It is important to understand how a binary star system forms from a molecular gas cloud because most main-sequence stars are in binary or multiple systems \citep[e.g.][]{abt83,duquennoy91}.
	Furthermore, recent observations indicate that companions may be even more common for pre-main sequence stars \citep{richichi94}.  
	Numerical simulations show that it is difficult for single stars to capture another one and to form a binary system in the main sequence phase \citep{kroupa01}.  
	Accordingly, most stars seem to be born as binaries rather than single stars in a gas cloud, and molecular bipolar outflows are often observed in star forming regions. 
	It has been shown using two-dimensional magnetohydrodynamical simulations that outflow plays an important role in the star formation process because the excess angular momentum, which is conserved from the host cloud, is removed by these outflows from the formed core \citep{tomisaka00, tomisaka02}.
	These outflows are caused by the twisted magnetic field lines made by the rotation near the adiabatic core.
	In order to investigate the cloud fragmentation and later binary star formation, three-dimensional calculation is required.
	In this letter, we discuss fragmentation in the course of star formation which leads to binary star formation.

      Fragmentation of isothermal cloud is studied by many authors using three-dimensional hydrodynamical simulations \citep[e.g.][]{bodenheimer00}.
        The criterion for an initially uniform cloud to fragment is described by $\alpha_0$ (thermal-to-gravitational energy ratio) of less than $\simeq 0.2-0.5$ if we restrict ourselves to the isothermal gas.
	The cloud with $\alpha_0 \ga 0.2 -0.5$ experiences the run-away collapse \citep{larson69,norman80}.
	When the gas density increases up to $n \, \ga \, 10^{10}$~cm$^{-3}$, the isothermal gas approximation breaks down.                                   
        Recently, \citet{bate02,bate03}, \citet{cha03} and \citet{matsu03b} have studied the cloud evolution from isothermal to adiabatic state using a barotropic equation of state and they clarified the evolution of the cloud with no magnetic field.
	However, there are few papers on the fragmentation of the magnetized cloud.
	\citet{boss02} studied the fragmentation taking account of radiative transfer with the Eddington approximation instead of barotropic equation of state, but simplified the effect of magnetic field.
	In his studies, the angular momentum transfer by the magnetic field is ignored.
	However, the angular momentum is largely removed by magnetic braking and molecular outflow in the cloud evolution process \citep{tomisaka00}.
	Therefore, to explore the fragmentation of the magnetized cloud, full three-dimensional magnetohydrodynamical MHD simulations are necessary.

	Here, we employ the nested grid MHD code, which always maintains sufficient spatial resolution in the central region.
	The nested grid method is powerful to study the star formation process, because this requires a large dynamic range in spatial dimensions.
	In the simulations, we use the barotropic equation of state instead of inclusion of radiative transfer.
	In this study, we calculate the non-axisymmetric dynamical contraction of the cloud from $4\times10^2$  cm$^{-3}$ to $1 \times10^{17}$  cm$^{-3}$ in number density, and  investigate how fragmentation proceeds and outflow occurs in the course of binary star formation.

\section{Model and Numerical Method}
	To study the star formation (binary star formation and their outflow) in the filamentary cloud \citep{mizuno95}
, we consider a cylindrical isothermal cloud in hydrostatic balance as the initial condition.
	The cloud is assumed to rotate parallel with a rotation axis which coincides with the cylindrical axis and the magnetic field lines are assumed to run parallel to the rotation axis, because magnetic field lines often turn to the cylindrical axis \citep{goodman90,ward00}.
 	Using the cylindrical coordinates ($r, \phi, z$), the density, azimuthal velocity, and magnetic flux density distributions in the radial direction are taken to be as follows \citep{sto63}: 
\begin{eqnarray}
\rho_0(r) &=& \rho_c  \left[ 1 + (r^2/8 H^2)  \right]^{-2},   \\
v_{\phi 0}(r) &=& r \Omega_c \left[ 1 + (r^2/8 H^2)  \right]^{-1/2},   \\
B_{z0}(r) &=& B_c \left[ 1 + (r^2/8 H^2)  \right]^{-1}, 
\end{eqnarray}
	where $\rho_c$, $\Omega_c$, and $B_c$ represent the density, angular rotation speed, and magnetic flux density at the center of the cylindrical cloud, respectively, and $H$ is the scale-height as $H^2 = (c_s^2 +B_c^2/8 \pi \rho_c)/(4 \pi G \rho_c  - 2 \Omega_c^2) ,$
	where $c_s$ denotes the isothermal sound speed.
	Molecular gas obeys the isothermal equation of state below $\rho_{\rm{cri}}\approx 10^{10} \ \rm{cm^{-3}}$ but it becomes adiabatic if  $\rho\ga\rho_{\rm{cri}}$ \citep{tohline82}.
	To mimic this, we adopt a two-components equation of state as 
$P= c_s^2 \rho + c_s^2 \rho_{\rm{cri}} \left( \rho/\rho_{\rm{cri}}
\right)^{7/5}$ \citep{bonnell94,bate02}.
	The initial central density and the critical density are chosen as $\rho_{c0} = 4\times10^2$  cm$^{-3}$ and $\rho_{\rm{cri}} = 10^{10}  \rm{cm}^{-3}$.

	To initiate contraction, we added axisymmetric and non-axisymmetric perturbations to the above magnetohydrostatic equilibrium as   
\begin{eqnarray}
\label{eq:1}
\rho(r, z, \phi)&=&\rho_0(r)[1+\delta\rho_z(z)][1+\delta\rho_{\phi}(r, \phi)], \\
B_z(r, z, \phi) &=& B_{z0}(r, \phi)[1+\delta B_{\phi}(r, \phi)], 
\label{eq:2}
\end{eqnarray}
with 
\begin{eqnarray}
\delta\rho_z(z)&=&A_z \cos(2\pi z/L),   \\
\delta\rho_{\phi}(r, \phi),\delta B_{\phi}(r, \phi) &=& \left\{ 
\begin{array}{ll}
\it{A_{\phi}}(r/H)^{\rm{2}} \cos(m\phi), \ \  \rm{for} \ \it{r_\le H},  \\
\it{A_{\phi}}\cos(m\phi), \ \ \ \ \ \  \rm{for} \ \it{r>H},
\end{array} 
\right.  
\end{eqnarray}
	where  $L$ is chosen equal to the wavelength of the most unstable Jeans mode acquired by the linear analysis \citep{matsu94}. 

	In this letter, we restrict ourselves to the relative amplitude of the axisymmetric perturbation $A_z=0.1$.
	As the non-axisymmetric perturbation, only $m$=2 mode is included.
	Equations~(\ref{eq:1}) and (\ref{eq:2}) ensure the ratio of the density to the magnetic flux constant in the azimuthal direction. 
	Models are parameterized with three non-dimensional parameters: the amplitude of the non-axisymmetric perturbation as $A_{\phi}$, the magnetic-to-thermal pressure ratio as $\alpha$($\equiv B^2_c/4\pi \rho_c c_s^2)$ and the angular speed normalized by the free-fall timescale as $\omega \equiv \Omega_c/(4\pi G \rho_c)^{1/2} $.
	We calculated 51 different models with $A_{\phi} = ( 0, \ 0.01, \ 0.1, \ 0.2 )$, 
$\alpha = ( 0, \ 0.01, \ 0.1, \ 1, \ 5 ),$ and $\omega = ( 0, \ 0.1, \ 0.5, \ 0.7 )$.  

	In order to solve the central region with higher spatial resolutions, we adopt an MHD nested grid method based on the Cartesian coordinate \citep[for detail, see][]{machida03}.
	In this method, a number of grids with different spacings are prepared, in which finer grids cover the central high-density portion and the coarser ones cover the cloud as a whole.
	Each grid has successively different cell widths by a factor two.
	We use $128 \times128 \times32$ cubic cells in the $x$-, $y$- and $z$-directions.
	We added a new finer grid to maintain the Jeans condition, $\lambda_J/4>h,$ with an ample margin \citep{truelove97}, where $\lambda_J$ and $h$ are the Jeans length and the cell width, respectively.
	Whenever one-eighth of the minimum Jeans length becomes smaller than the cell width of the finest grid, a new finer grid is added to the nested grid. 
      Total number of cells is $128\times128\times32\times \ell_{\rm max},$ where $\ell_{\rm max}(\lid 17)$ denotes the maximum grid level.
      Only when a high density fragment escapes from the region covered by the finest grid, the Jeans condition is violated in our simulations.

	The MHD code uses the numerical fluxes proposed by \citet{fukuda99} but modified to solve the isothermal and polytropic gas.
We adopt the MUSCL approach and predictor-corrector method in the time integration.
The Poisson equation is solved by the multigrid iteration method on the nested grid \citep{matsu03a}.
  	We adopt mirror symmetry with respect to the $z=0$ and $z = L/2$ planes.
	We also set the fixed boundary condition on the surface of $r=2H$.

\section{Results}
\label{sec:results}
	The cylindrical gas cloud collapses to become spherical in shape in the early collapse phase. 
	If the initial gas cloud has neither magnetic field nor rotation speed, the cloud continues to collapse spherically and it forms into a small massive core in the center of the cloud \citep{larson69} even in a cylindrical cloud \citep{tomisaka95}.
	In the case of a gas cloud with magnetic fields or rotation, it collapses to form a pseudo-disk due to the magnetic pressure or centrifugal force \citep{tomisaka95,nakamura97,matsu97}. 
	If only the axisymmetric perturbation exists, a round disk is formed, but it does not fragment as long as the axisymmetry holds.
	When the non-axisymmetric perturbation is added, the cloud evolves to form a non-axisymmetric shape, and fragments later for some range of the parameters. 
In any case, the central density ($\rho_c$) increases with time and the gas becomes adiabatic when $\rho_c >10^{10}\ \rm{cm^{-3}}$.
	For convenience, we divide the evolution into two phases: collapse phase ($\rho_c<10^{10} \rm{cm^{-3}}$) and accretion phase ($\rho_c>10^{10} \rm{cm^{-3}}$).
	In this equation of state, the temperature gradually increases near $\rho \la \rho_{\rm cri}$, however, the solution acquired in $\rho<\rho_{\rm cri}$ is almost the same as that obtained for the isothermal collapse.
The outflow is driven by the twisted magnetic fields \citep{tomisaka98,tomisaka02}
 after $\sim 10^3$ yr have passed from the core formation epoch $\rho_c=\rho_{\rm{cri}}$.

\begin{figure}
\includegraphics[width=160mm]{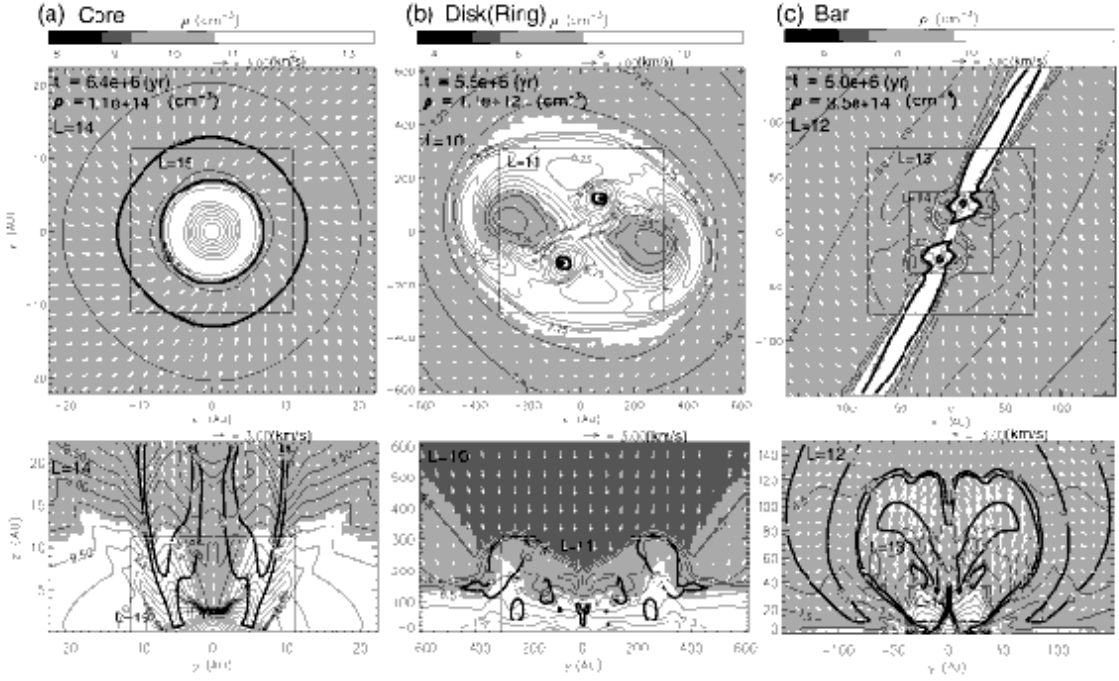}
\caption{
Three typical models of the adiabatic core are plotted (core-type: left, disk- or ring-type: middle, bar-type: right).
Model parameters are ($A_{\phi}, \alpha, \omega$) = (a): (0.0, 0.1, 0.1) , (b):(0.01, 0.01, 0.5), and  (c):(0.2, 1.0, 0.5).
 Density (contour and false color) and velocity vectors (arrows) are plotted on a plane including the major axis of the core and the $z$-axis (lower panels) and on the $z$=0 plane (upper panels). The elapsed time, the maximum density, grid level and scale length are also plotted.
Thick lines in the upper panels denote the contour line of $\rho=10^{10} \rm{cm^{-3}}$ showing the outline of the adiabatic core, while those in the lower panels mean the isovelocity curves representing the outflow region ($v_{\rm{z}}=0.1, 1, $ and $4 \ \rm{km \ s^{-1}}$)}
\label{fig:1}
\end{figure}

	Fig.~\ref{fig:1} shows the pole-on views  of the adiabatic cores (upper panels) and  the side-views of the outflow regions (lower panels) for three models in the accretion phase.
	The thick line denotes the boundary of the adiabatic core ($\rho\geq\rho_{\rm{cri}}$) for the upper panels.
	In the lower panels, the thick lines which represent the isovelocity lines of $v_z=0.1, 1$ and $4 \ \rm{km \ s^{-1}}$ indicate the outflow region.
	In Fig.~\ref{fig:1}{\it a}, a model without non-axisymmetric perturbations ($A_{\phi} = 0$) is shown.
	Model parameters $(\alpha, \omega) = (0.1, 0.1)$ correspond to the cloud with a weak magnetic field and slow rotation.
	This cloud collapses almost spherically and forms a small spherical core at the center of the cloud.
	We have no evidence of fragmentation in this model.
On the other hand, in the studies of \citet{boss02} and \citet{li02}, fragmentation occurs in this parameter.
	However, their results are not inconsistent to ours, taking account of the fact that the magnetic braking is ignored in their calculation.
	We also found a strong outflow is driven near the adiabatic core after the central density reaches $\sim10^{13} \ \rm{cm}^{-3}$ (lower panel).

	A model with small non-axisymmetric perturbation ($A_{\phi}=0.01$) is shown in the middle panel (b):($\alpha, \omega) = (0.01, 0.5)$. 
	In this model, a pseudo-disk is formed in the isothermal collapse phase, because the gas cloud is rotating rapidly ($\omega=0.5$).
	The disk evolves into a ring in the accretion phase, and fragmentation occurs on the inner edge of the ring. 
	Two contracting cores at $(x,y)$ = ($75 \  \rm{AU}, 125 \ \rm{AU}$) and ($-75 \ \rm{AU}, -125 \ \rm{AU}$)  are formed.
	The right panel (c):$(\alpha,\omega)=(1,0.5)$ shows the model with a large amplitude of non-axisymmetric perturbations ($A_{\phi}=0.2$).
	In this case, non-axisymmetric patterns grow sufficiently in the isothermal collapse phase and a bar is formed at the beginning of the accretion phase.
	In the accretion phase, the bar grows longer and thinner.
	The bar fragments and forms two contracting cores [$(x, y)$=$(-10 \ \rm{AU}, -25\ \rm{AU}), (10 \ \rm{AU}, 25 \ \rm{AU})$].
	Fig.~\ref{fig:1} shows that there are at least two kinds of fragmentation: fragmentation from a ring (b) and  from a bar (c).

        Fig.~\ref{fig:1}{\it c} shows that lobes of outflow are ejected from each contracting core ($y\simeq\pm 25 \rm{AU}$), whose velocities reach $\simeq$8 km s$^{-1}$.
	As for the outflow, low-velocity outflows are seen also in Fig.~\ref{fig:1}{\it b}, each of which is connected to the contracting core.
	In both the bar and ring fragmentations, two types of outflows co-exist: the inner one is connected to the cores formed by the fragmentation and driven by their spin motion, and the other is driven by the rotation motion of the ring or the bar.
	However, the strength of the outflow is different for respective fragmentation patterns.
	Fast outflows appear in the bar fragmentation (c), while weak ones appear in the ring fragmentation (b).
	This difference comes from the angular momentum distribution between the spin motion of the cores and their orbital angular momentum.
	In the ring fragmentation, the fragments have the angular momentum in the form of orbital angular momentum.
	On the other hand, in the bar fragmentation, the angular momentum is almost evenly shared between the spin and the orbital angular momenta.
	In the model of Fig.~\ref{fig:1}{\it c}, the fragments are expected to merge each other due to their small orbital angular momentum.\\

\section{Discussion}
	As mentioned in \S~\ref{sec:results}, modes of fragmentation and shapes of the outflow depend on the growth of non-axisymmetry.
	This shows that the cloud evolution can be classified by this growth.
	In order to characterize the shape of the central region, we define oblateness ($\varepsilon_{\rm{ob}} $) and axis ratio ($\varepsilon_{\rm{ar}} $) as follows:
\begin{eqnarray}
\varepsilon_{\rm{ob}}  &\equiv& \sqrt{h_l\cdot h_s }/h_z , \\
\varepsilon_{\rm{ar}}  &\equiv& h_l/h_s ,
\end{eqnarray}
where $h_l$, $h_s$ and $h_z$ mean the length of the long-, short-, and $z$-axis for the gas with $\rho>(1/10) \rho_{\rm c}$, respectively \citep[see][for definition of $h$]{matsu99}. 

\begin{figure}
\includegraphics[width=160mm]{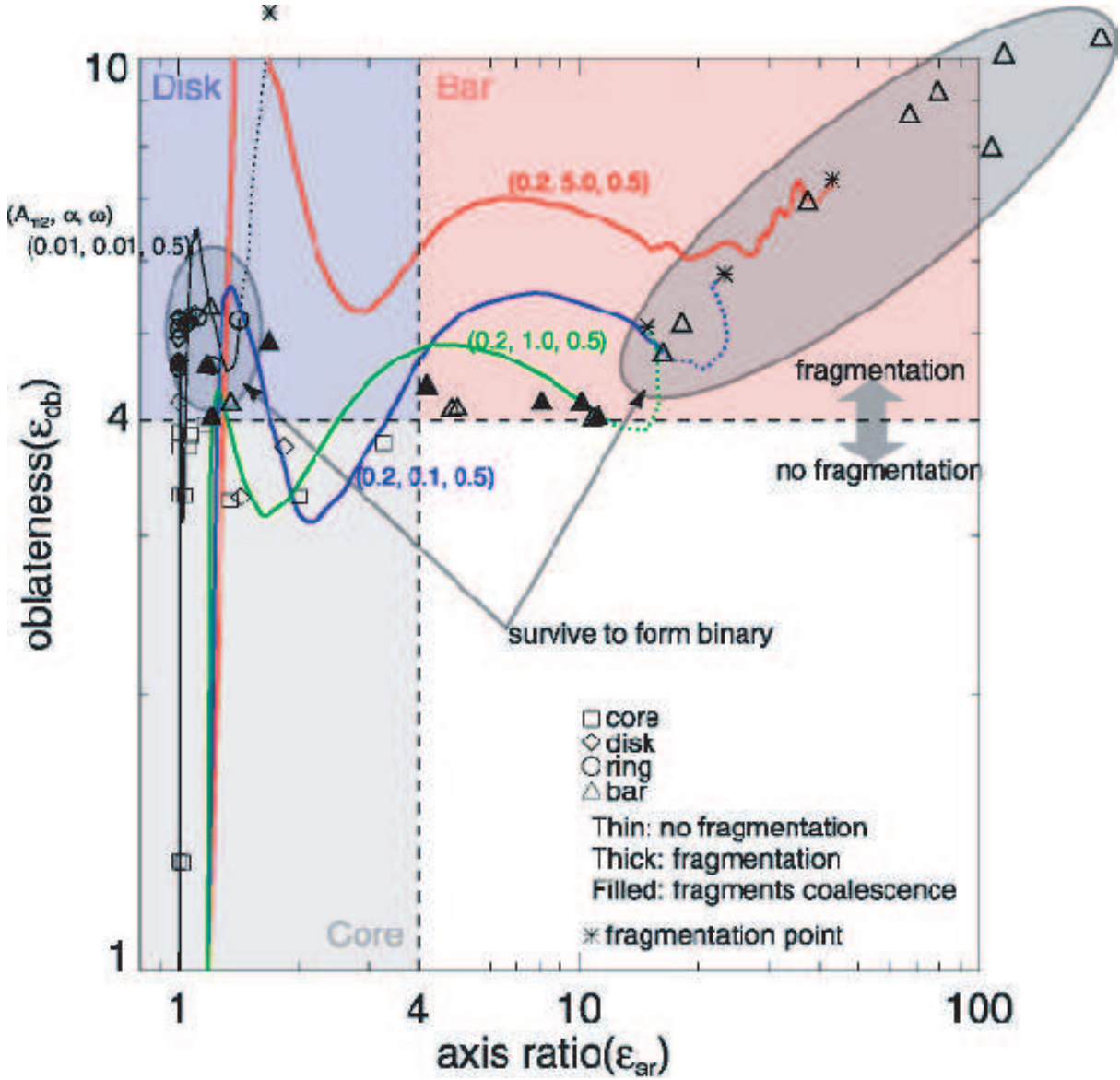}
\caption{
The axis ratio and oblateness are plotted when the gas at the cloud center becomes adiabatic $\rho_c=\rho_{\rm{cri}}$.
The abscissa indicates the axis ratio and the ordinate is the oblateness.
Both are calculated for the gas with $\rho>(1/10)\rho_{\rm c}$.
Fragmentation occurs above the horizontal dashed line ($\varepsilon_{\rm ob}>4$).
The domains in large circles mean that fragmented cores survive to form binary or multiple stars.
The solid lines represent the evolutional tracks for some typical parameters.
Dotted lines represent the evolution after the adiabatic core is formed.
Asterisks show the points of fragmentation.
The symbols $\sq$ (core), $\diamondsuit$ (disk), $\bigcirc$ (ring) and $\bigtriangleup$ (bar) denote the shape of the adiabatic core when the fragmentation occurs or the calculation ends.
Thick-line and filled symbols represent the cores which experience fragmentation.
Thin-line symbols represent the non-fragmentation models.
However, filled symbols indicate merger core after fragmentation.
 }
\label{fig:2}
\end{figure}

	In Fig.~\ref{fig:2}, the oblateness and the axis ratio of the central region are plotted for all models at the beginning of the accretion phase ($\rho_c =\rho_{\rm{cri}}=10^{10} \ \rm{cm^{-3}}$). 
	Curves in this figure illustrate the evolutionary tracks for four different models ($A_{\phi}, \alpha, \omega$)= (0.01, 0.01, 0.5),  (0.2, 0.1, 0.5), (0.2, 1, 0.5), and (0.2, 5, 0.5). 
	The clouds evolve from the lower-left corner, which means the early shape is almost spherical, to the upper-right region via the upper-left region. 
	First, the clouds evolve vertically upward.
	This indicates that the oblateness increases.
	Then, the curves change their directions to the right for three models with $A_{\phi} =0.2$.
	Evolutionary tracks of the high-density region in
	Fig.~\ref{fig:2} show very clearly that the non-axisymmetric perturbation grows only after the disk is formed.
	That is, the axis ratio increases after the oblateness grows considerably large ($\varepsilon_{\rm{\rm{ob}} } \ge4$).
	Because the gas is partially supported in the radial direction by the magnetic pressure and the centrifugal force, the radial contraction is more delayed than that of the $z$-direction.
	As a result, the disk is formed ($\varepsilon_{\rm{ob}}  > 4$) earlier for models with larger $\alpha$ and $\omega$.
	In such models, the non-axisymmetry grows sufficiently and forms a large bar in the isothermal collapse phase. 
	However, the non-axisymmetry hardly evolves in the model with small $\alpha$ and $\omega$, because it takes a long time for disk formation.

	Adiabatic cores are classified into three types: $core$ (Fig.\ref{fig:1}a: $\varepsilon_{\rm{ob}} <4$ and $\varepsilon_{\rm{ar}} <4$), $disk$ (Fig.\ref{fig:1}b$:  \varepsilon_{\rm{ob}} >4$  and $\varepsilon_{\rm{ar}} <4$), and $bar$ (Fig.\ref{fig:1}c: $ \varepsilon_{\rm{ob}} >4$ and $\varepsilon_{\rm{ar}} >4$).
	Some $disks$ evolve into the rings, in which the central density is lower than that outside, even if it has the same oblateness and the axis ratio.   
	Models are divided into three kinds of evolution: thin-line symbols mean that the models form cores which continue to collapse, while thick-line and filled symbols represent the models in which fragmentation occurs.
        Although the fragments survive against mutual merger in the models of 
thick-line symbols, they merge each other in the models of filled symbols.

	Since the models with thick-line or filled symbol are distributed only in the region $\varepsilon_{\rm{ob}} >4$, it is concluded that the necessary condition for fragmentation is $\varepsilon_{\rm{ob}} >4$, which means that a thin disk or bar is necessary for fragmentation.
	Even in the region of $\varepsilon_{\rm ob}>4$, there are exceptional three models in which the adiabatic core evolves to form an almost axisymmetric thin disk or ring because non-axisymmetry is counteracted in the isothermal collapse phase.
	The shapes of the $bar$ and $ring$ are also seen in \citet{boss02} and \citet{li02}.
	It is concluded that for the adiabatic core to fragment, the central gas needs to form a sufficiently thin disk or bar in the isothermal collapse phase.
	The thick-line symbols are distributed in two regions enclosed by circles.
	The fragmented cores in the circles seem to evolve into a binary star because the length of the semi-major axis expands or oscillates.
	Models in which the fragments merge with each other are distributed outside the domain.
	Outside of the domain, a few fragments survive at the end of the calculation but the separation between the two cores shrinks with time.
	These will not grow into binary stars. 

	In conclusion, if the following criteria are fulfilled at the end of the isothermal collapse phase, fragmentation occurs and fragments survive to form binary or multiple stars.
\begin{enumerate}
\item $\varepsilon_{\rm{ob}} >4$ : fragmentation condition.
\item $0<\varepsilon_{\rm{ar}} <2$, or  $\varepsilon_{\rm{ar}} >10$ :
	survival condition against merger. 
\end{enumerate} 
	The symbol in this figure indicates the shape of the adiabatic core when the fragmentation occurs or the calculation ends. 
	Fig.~\ref{fig:2} confirms the idea that the fragmentation occurs from a ring 
($\bigcirc$) or bar ($\bigtriangleup$) and not from a core ($\sq$).

	We showed the domain which may evolve into binary stars in Fig.~\ref{fig:2}.
	Outside the binary-forming domain, the core formation model ($\sq$) results in a compact massive core at the center of cloud.
	The bar-fragmentation model ($\bigtriangleup$) in which the mutual merger of the fragments occurred also leads to a disk-like core at the center of cloud.
This configuration seems to lead to single star formation.

\section*{Acknowledgments}
We have greatly benefited from discussion with Prof.~ T.~ Hanawa  and Dr.~ H.~ Koyama.
Numerical calculations were carried out by Fujitsu VPP5000 at the Astronomical Data Analysis Center, the National Astronomical Observatory of Japan.
This work was supported partially by the Grants-in-Aid from MEXT (11640231, 14540233 [KT], 14740134[TM]).

\clearpage

\clearpage

\end{document}